\documentclass[twocolumn]{revtex4-2}
\usepackage[utf8]{inputenc}

\usepackage{amsmath} 
\usepackage{amsfonts}
\usepackage{amssymb}
\usepackage{bbm} 

\usepackage{graphicx}
\graphicspath{ {./figures/} }

\newcommand{\minMag}{\ensuremath\beta}
\newcommand{\quality}{\ensuremath{q_0}}
\newcommand{\numMags}{\ensuremath{n}}

\begin{document}

\title{Assessing the quality of a network of vector-field sensors}
\author{Joseph A. Smiga}
\email{jsmiga@uni-mainz.de}
\affiliation{Helmholtz-Institut Mainz, GSI Helmholtzzentrum f{\"u}r Schwerionenforschung, 64291 Darmstadt, Germany}
\affiliation{Johannes Gutenberg-Universit\"at Mainz, 55128 Mainz, Germany}

\begin{abstract}
    An experiment consisting of a network of sensors can endow several advantages over an experiment with a single sensor: improved sensitivity, error corrections, spatial resolution, etc. However, there is often a question of how to optimally set up the network to yield the best results. Here, we consider a network of devices that measure a vector field along a given axis; namely for magnetometers in the Global Network of Optical Magnetometers for Exotic physics searches (GNOME). We quantify how well the network is arranged, explore characteristics and examples of ideal networks, and characterize the optimal configuration for GNOME. We find that by re-orienting the sensitive axes of existing magnetometers, the sensitivity of the network can be improved by around a factor of two. 
\end{abstract}
\maketitle

\section{Introduction}




Various experiments make use of a network of sensors in lieu of a single, centralized device. A network of sensors can come with several advantages such as having better sensitivity than a single device, being able to catch and correct errors, and achieving superior spatial resolution. However, there are also a few challenges and complexities that arise when involving many devices. In addition to the logistical challenges of managing multiple devices at once and making sense of several data streams, there is the foundational question of how to best arrange the network. This question is explored for a network consisting of a specific class of sensors: ones that measure a vector field. 

A common motivational interest in designing these network experiments is in measuring spatially extended phenomena. For example, interferometer networks used to measure gravitational waves~\cite{anderson_excess_2001, adhikari_gravitational_2014, klimenko_method_2016} and gravimeters used for geodesy~\cite{boy_achievements_2020}, and magnetometers used for geophysics~\cite{gjerloev_global_2009, gjerloev_supermag_2012, bergin_ae_2020}; all of which measure vector-field phenomena\footnote{The interferometer networks can be understood as measuring something closer to a tensorial deformation in spacetime. However, similar to vector-field sensors that measure a vector only along one axis, these interferometers are only sensitive to certain polarizations of gravitational waves.} on scales the size of the Earth or larger. 
Networks have also been used to search for direct evidence of dark matter, which is believed to dominate the mass of the galaxy but only weakly interacts with visible matter; see, e.g., reviews in Refs.~\cite{feng_dark_2010, gorenstein_astronomical_2014, marsh_axion_2016}. This includes both gravimeters~\cite{horowitz_gravimeter_2020, hu_network_2020, mcnally_constraining_2020, figueroa_dark_2021} that search for the motion of dark matter captured by the Earth as well as magnetometers~\cite{pospelov_detecting_2013, jackson_kimball_searching_2018, masia-roig_analysis_2020, fedderke_earth_2021, fedderke_search_2021} that search for coupling between dark and visible matter.

For this work, the Global Network of Optical Magnetometers for Exotic physics searches (GNOME)~\cite{pospelov_detecting_2013, jackson_kimball_searching_2018, masia-roig_analysis_2020} is of particular interest. GNOME consists of shielded magnetometers around the Earth and has the goal of finding new, exotic (vector) fields that couple to fermionic spin. For example, GNOME searches for axion-like particle (ALP) domain walls~\cite{pospelov_detecting_2013, masia-roig_analysis_2020} via the coupling of the ALP field gradient to nucleon spin. The gradient, in this case, is in effect a vector field; albeit with typical constraints, such as having a vanishing curl. 

This paper is organized as follows: a method of calculating the network sensitivity is described in Sec.~\ref{sec:sensitivity}, a quantification of network quality is described in Sec.~\ref{sec:qualFactor}, ideal and optimized networks are described in Sec.~\ref{sec:idealOptNet}, and concluding remarks are given in Sec.~\ref{sec:conclusion}. Throughout this paper, the network under consideration will consist of the GNOME magnetometers. However, the principles explored here can be extended to other network experiments.

\section{Sensitivity}\label{sec:sensitivity}

The magnetometers in the network each possess a ``sensitive axis'' that result in the attenuation of a signal when the vector field is not parallel or anti-parallel to the sensitive axis. Denote the sensitive axis of magnetometer $i$ with $\boldsymbol{d_i}$. The magnitude of this vector reflects the strength of the coupling such that a vector (field) $\boldsymbol{m}$ will induce a signal $s_i=\boldsymbol{d_i}\cdot\boldsymbol{m}$ in the $i^\text{th}$ magnetometer. Consider the case in which only one vector $\boldsymbol{m}$ describes the signal. For a domain wall, this could be the gradient at the center of the wall with the timing of the signal adjusted to account for delays as the domain wall crosses the network. The signals observed by the network can be simplified into the linear equation,
\begin{equation}\label{eq:signalLinEq}
D\boldsymbol{m} = \boldsymbol{s}\,,
\end{equation}
where $D$ is a matrix whose rows are $\{\boldsymbol{d_i}\}$ and similarly $\boldsymbol{s}=\{ s_i \}$. 

In practice, one will measure a set of signals $\boldsymbol{s}$ with some error. Let $\Sigma_s$ be the covariance matrix in the measurements $\boldsymbol{s}$ that characterizes this error; the matrix will generally be diagonal as noise between the GNOME magnetometers is uncorrelated. By $\chi^2$ minimization, Eq.~\eqref{eq:signalLinEq} is approximately solved by
\begin{subequations}
\begin{align}
\boldsymbol{m} &= \Sigma_m D^T \Sigma_s^{-1} \boldsymbol{s} \\
\textrm{for}\quad \Sigma_m &= \left( D^T \Sigma_s^{-1} D \right)^{-1}\,,
\end{align}
\end{subequations}
where $\Sigma_m$ is the covariance matrix for $\boldsymbol{m}$. 

The network sensitivity is defined as the magnitude necessary to induce a $\lVert\boldsymbol{m}\rVert$ signal-to-noise ratio $\zeta=\lVert\boldsymbol{m}\rVert/\sqrt{\hat{\boldsymbol{m}}^T \Sigma_m \hat{\boldsymbol{m}}}$ of one. Thus, the sensitivity in the direction $\hat{\boldsymbol{m}}$ is
\begin{equation}\label{eq:sensitivity}
\minMag(\hat{\boldsymbol{m}}) = \sqrt{\hat{\boldsymbol{m}}^T \left( D^T \Sigma_s^{-1} D \right)^{-1} \hat{\boldsymbol{m}}}\,.
\end{equation}
This will vary by direction. However, one can find the range of sensitivities over different directions by solving for the eigenvalues of the symmetric, positive-definite matrix $\Sigma_m^{-1} = \left( D^T \Sigma_s^{-1} D \right)$ --- the smallest eigenvalue $\lambda_\text{min} = \minMag_0^{-2}$ giving the ``worst-case'' sensitivity as a large signal $\minMag_0$ would be needed to induce a significant signal. Likewise the largest eigenvalue $\lambda_\text{max} = \minMag_1^{-2}$ gives the ``best-case,'' and the corresponding eigenvectors are the directions that induce such signals.

\section{Quality factor}\label{sec:qualFactor}
Given the sensitivity defined by Eq.~\eqref{eq:sensitivity}, there remains the question of how to optimize the network; in particular, how to optimize the directions $\{\boldsymbol{d_i}\}$ for the best network. If there is distribution of directions of interest for the vector field, one could define the optimization by performing some weighted average of Eq.~\eqref{eq:sensitivity} over this distribution. If one is ambivalent about the direction of the signal, the worst-case direction indicates a bound on sensitivity.

Ideally, the magnetometers in the network will be oriented to evenly cover all directions. If there is a preferred and unpreferred direction, one could improve the sensitivity in the unpreferred direction by rotating the sensitive axis of magnetometers towards this direction. Under practical conditions, it is not possible to have GNOME always operating under optimal conditions because the noise in individual sensors varies over time and magnetometers will occasionally activate and deactivate. 

\begin{figure*}[ht]
	\centering
	\includegraphics[width=\textwidth]{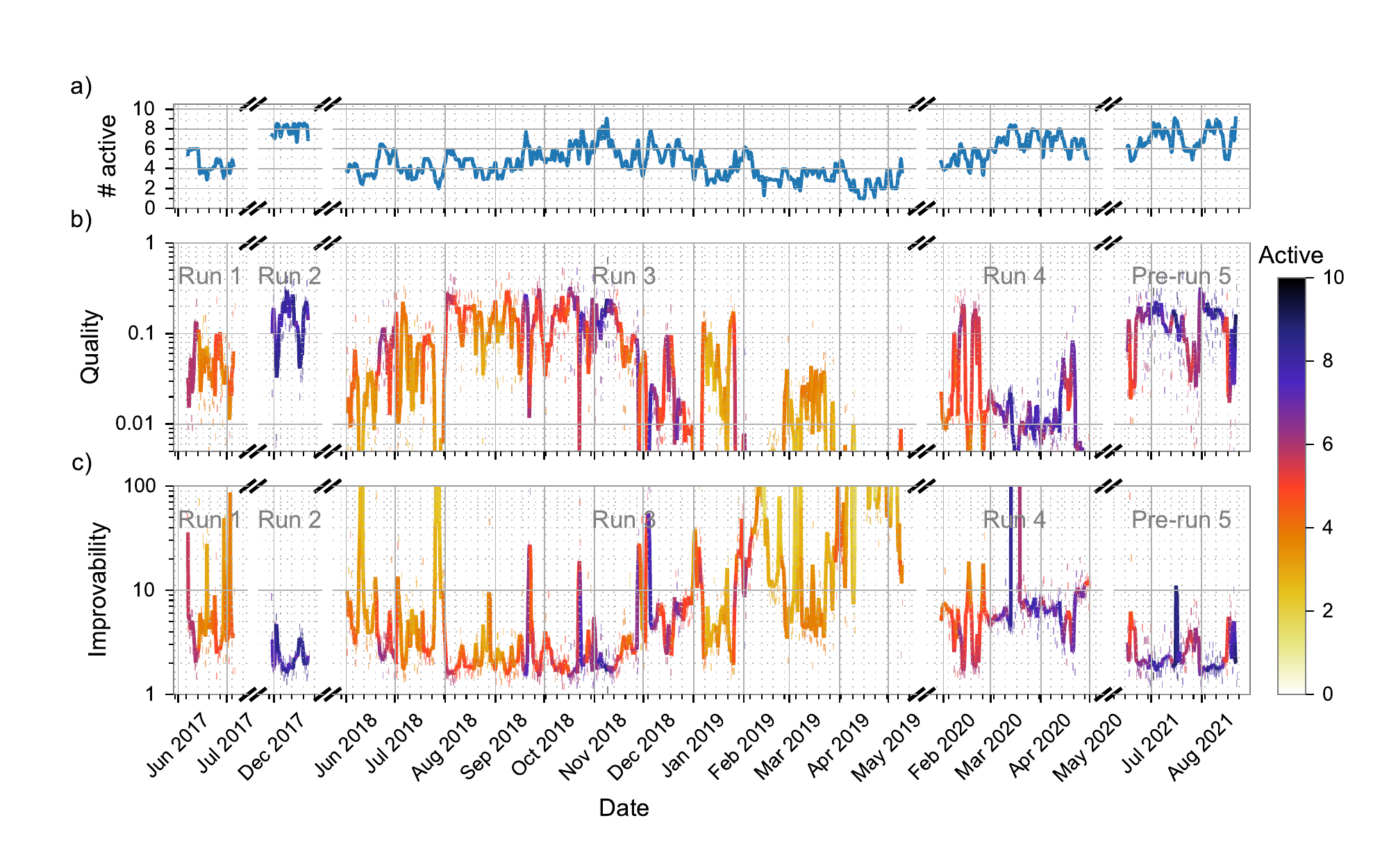}
	\caption{The quality factor over time for Science Runs~1--(pre)5. Solid lines represent 1~day rolling averages. (a) The number of active sensors over time. (b) The quality factors over time. The color indicates the number of active sensors. (c) The approximate factor by which the network could be improved if the sensors were optimally oriented using the approximate optimized sensitivity, Eq.~\eqref{eq:optSensitive}.}
	\label{fig:qualOverTime}
\end{figure*}

To judge how well the GNOME network is performing, it helps to define some quantitative ``quality factor.'' This factor would ideally reflect how optimally the network is set up with the magnetometers available and not the absolute sensitivity of the network. That is, the quality of the network refers to how well the magnetometers are oriented and is not affected by improving all magnetometers by a constant factor. One possibility is the quotient of the best and worst sensitivity,
\begin{equation}\label{eq:qualityFactor}
\quality := \minMag_1/\minMag_0\,.
\end{equation}
This factor will be between zero and one with a more optimally oriented network having a larger value. The quality factor for GNOME during the Science Runs is given in Fig.~\ref{fig:qualOverTime}b.

A few terms will be defined based on the quality of a network. Namely, an ``ideal'' network is one for which $\quality=1$, while an ``optimal'' network is one in which the sensitivity $\minMag_0$ cannot be improved by re-orienting the sensors in the network. If the best sensitivity is the one that optimizes sensitivity in the least-sensitive direction, an ideal network will also be optimal.

Given a set of magnetometers with covariance matrix $\Sigma_s$ and known coupling, it should be possible to determine a theoretical best sensitivity. For this, we will consider a network with $\numMags$ independent magnetometers that are all described by a single sensitive axis $\boldsymbol{d}_i$ with coupling strength $\kappa_i := \lVert \boldsymbol{d}_i \rVert$. 
Consider the case in which the angle between any vector signal and any sensitive axis is random (this would be the case for many randomly oriented sensors). Because $\left( \hat{\boldsymbol{d_i}} \cdot \hat{\boldsymbol{m}} \right)^2 \approx \left< \cos^2\theta \right> = 1/d$, for $d=3$ spatial dimensions, Eq.~\eqref{eq:sensitivity} in this case becomes,
\begin{equation}\label{eq:optSensitive}
\minMag_\text{opt} \approx \sqrt{ d/\sum_{i=0}^{\numMags-1} \left( \kappa_i^2/\sigma_i^2 \right) }\,.
\end{equation}
This may not be the optimal sensitivity, but it provides a heuristic for an optimal network. The quotient between the observed and optimal sensitivity for GNOME over time is given in Fig.~\ref{fig:qualOverTime}c. 

\begin{figure*}
	\centering
	\includegraphics[width=\textwidth]{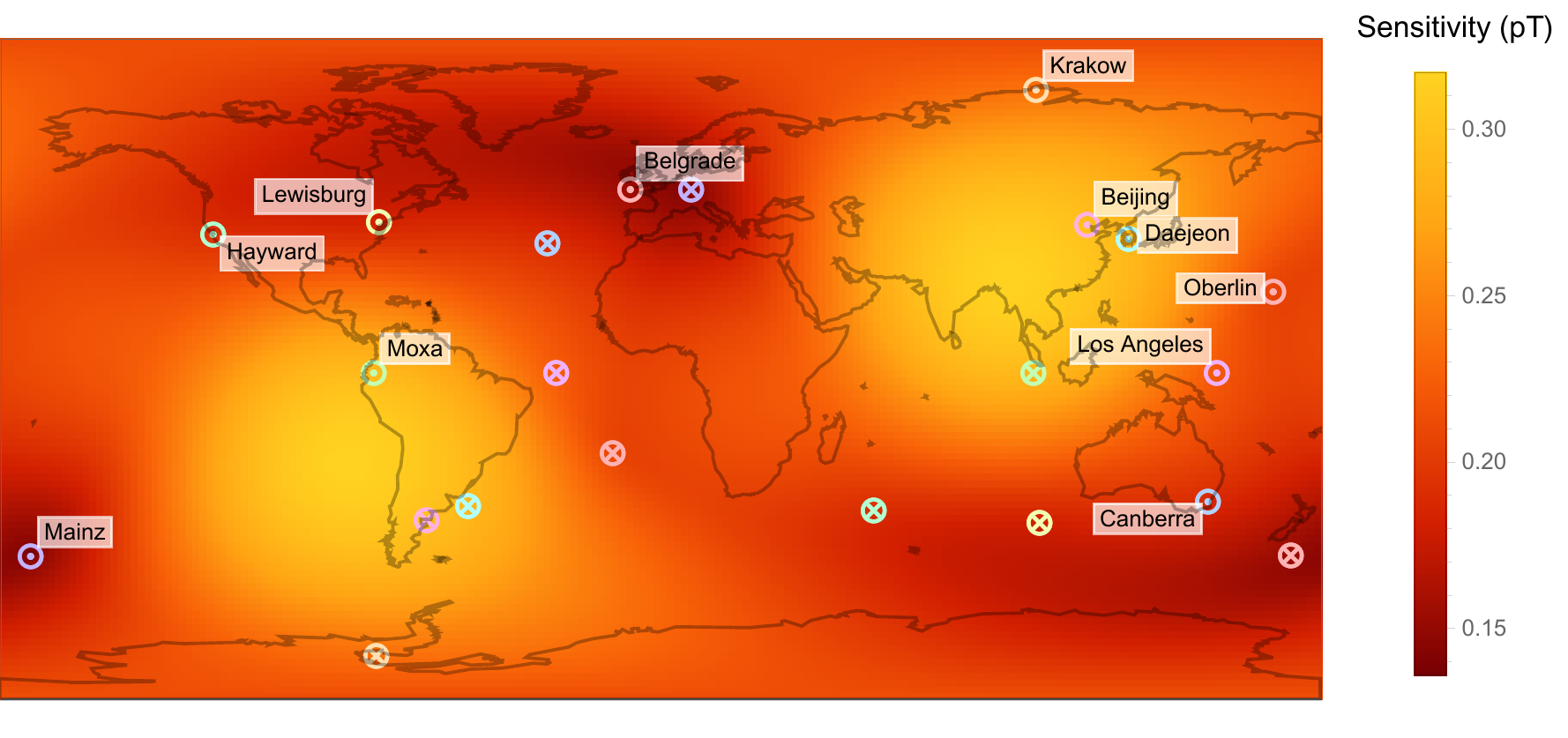}
	\caption{Average sensitivity $\minMag$ of GNOME during Science pre-Run~5 (16~June--23~August 2021). A position on the Earth represents the direction perpendicular to that point on the Earth. The sensitive axes of the stations are represented by $\odot$ (parallel) and $\otimes$ (anti-parallel). The marker color is a visual aid to associate the pair of markers representing the same station.}
	\label{fig:sensMap}
\end{figure*}

With the quality factor in mind, it helps to consider exactly how sensitivity varies with direction. The network has been fairly stable with many active sensors during the recent pre-run for Science Run~5. A map of the average sensitivity, $\left< \minMag\left( \hat{\boldsymbol{m}} \right)^{-1}\right>^{-1}$, in different directions is shown in Fig.~\ref{fig:sensMap}. The network quality could be improved by improving the quality/reliability of stations sensitive to insensitive directions (e.g., Beijing or Daejeon) or rotating/adding additional sensor(s) towards the worst direction.

\section{Ideal and optimized networks}\label{sec:idealOptNet}
With the quantitative definition of network quality given in the previous section, various optimized and ideal networks will be given here. In particular, we consider properties and explicit arrangements of ideal networks as well as numerical optimizations of more realistic networks.

To better understand the characteristics of networks, it helps to define some additional formalism. Define a network with a given set of orientations as a pair of directional matrix and covariance matrix $N=\{D,\Sigma\}$. Two networks $N_0 = \{D_0,\Sigma_0\}$ and $N_1 = \{D_1,\Sigma_1\}$ can be considered equivalent $N_0\cong N_1$ if there exists a permutation matrix $P$ such that $D_1 = PD_0$ and $\Sigma_1 = P\Sigma_0P^T$; that is, they are the same up to ordering. Additionally, a network can be decomposed into two complementary subnetworks $N\cong N_A\oplus N_B$ if
\[
N\cong \left\{ \left[ \begin{matrix} D_A \\ D_B \end{matrix} \right], \left[ \begin{matrix}\Sigma_A & 0 \\ 0 & \Sigma_B \end{matrix} \right] \right\}\,.
\]
Observe that the subnetworks $N_A$ and $N_B$ are independent/uncorrelated. 

In addition to the basic equivalence relation described above, there are some additional symmetries for a network. First, the sensitivity $\minMag(\hat{\boldsymbol{m}})$ is invariant with respect to parity reversal of a subnetwork; i.e., $D_A\mapsto - D_A$ for $N\cong N_A\oplus N_B$. Further, though the sensitivity $\minMag(\hat{\boldsymbol{m}})$ of a (non-ideal) network can change under arbitrary rotation of the whole network $D^T\mapsto RD^T$, the value of the worst sensitivity does not. 

\subsection{Ideal networks}
There are a few useful characteristics of ideal networks that are worth considering. For an ideal network, $\quality=1$ so the smallest and largest eigenvalues of $D^T\Sigma^{-1} D$ are the same which implies that this matrix is proportional to the identity matrix. In particular, $D^T\Sigma^{-1} D = \minMag^{-2}\mathbbm{1}$. It also follows that the sensitivity of an ideal network is independent of global rotations.

Consider, now, how ideal networks are combined. Let $N_A$ and $N_B$ be two complementary, ideal subnetworks of $N$, then
\[
\left[\begin{matrix}{D_A}^T & {D_A}^T\end{matrix}\right]
\left[\begin{matrix}\Sigma_A & 0 \\ 0 & \Sigma_B \end{matrix}\right]^{-1}
\left[\begin{matrix} D_A \\ D_B\end{matrix}\right] = \left( \minMag_A^{-2} + \minMag_B^{-2} \right)\mathbbm{1}\,.
\]
That is, the network $N\cong N_A \oplus N_B$ is also ideal with sensitivity $\left( \minMag_A^{-2} + \minMag_B^{-2} \right)^{-1/2}$. Further, because an ideal network remains ideal under rotations, one can rotate either ideal subnetwork without affecting the sensitivity of the network $N$. Further, if $N=N_A\oplus N_B$ is an ideal network and $N_A$ is an ideal subnetwork, then $N_B$ is also an ideal subnetwork. An ideal network that cannot be separated into ideal subnetworks is ``irreducible.'' Because an ideal network needs at least $d$ sensors (for $d=3$ spatial dimensions), and ideal network with $n<2d$ is irreducible, because it cannot be split into two ideal networks.

\begin{table}
    \newlength{\cellH}
    \setlength{\cellH}{50pt} 
    \centering
    \caption{Examples of optimal networks based on the platonic solids. The orientations of sensitive axes in an ideal network are given as lines (dashed in one direction, solid in the other). The number of vertices are separated by ideal network. For example, the cube has eight vertices, and the corresponding ideal network consists of two ideal networks with four sensors each; hence ``$4+4$'' is listed. In each case here, $X+X$ vertices describes two ideal networks with $X$ sensitive axes in opposite directions.}
    \begin{tabular}{lll}
        \hline\hline
        Name & Vertices & Shape \\
        \hline
        \raisebox{.5\cellH}{Octahedron} & \raisebox{.5\cellH}{$3+3$} & \includegraphics[width=\cellH]{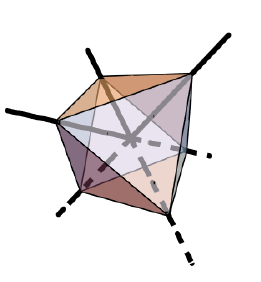} \\
        \raisebox{.5\cellH}{Tetrahedron} & \raisebox{.5\cellH}{$4$}  & \includegraphics[width=\cellH]{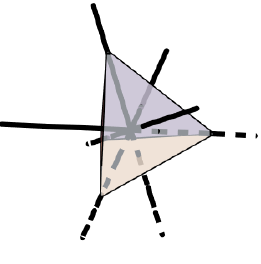} \\
        \raisebox{.5\cellH}{Cube} & \raisebox{.5\cellH}{$4+4$} & \includegraphics[width=\cellH]{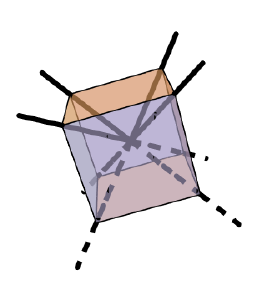} \\
        \raisebox{.5\cellH}{Icosahedron} & \raisebox{.5\cellH}{$6+6$} & \includegraphics[width=\cellH]{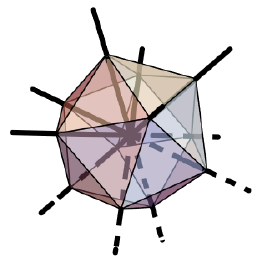} \\
        \raisebox{.5\cellH}{Dodecahedron} & \raisebox{.5\cellH}{$10+10$} & \includegraphics[width=\cellH]{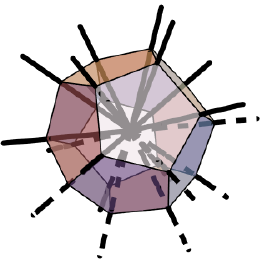} \\
        \hline\hline
    \end{tabular}
    \label{tab:optNetworks}
\end{table}

One can consider certain explicit cases of ideal networks with some simplified conditions. In particular, let $N=\{ D, \Sigma \}$ be composed of $n$ identical, independent sensors --- that is, $\Sigma = \sigma^2\mathbbm{1}$ and $\lVert\boldsymbol{d}_i\rVert = \kappa$ where $\boldsymbol{d}_i$ is the $i^\text{th}$ row of $D$. If there are $d=3$ sensors oriented such that their sensitive axes are orthogonal, then the resulting network will be ideal with sensitivity $\minMag=\sigma/\kappa$. 
Heuristically, one would like to orient the magnetometers to evenly in all directions. One way to do this is to take some inspiration from the Platonic solids by designing a network in which the sensitive axes of the sensors are oriented from the center of the solid to each of the vertices; see Table~\ref{tab:optNetworks}. Most of these solids will generate a network with two ideal subnetworks having opposite sensitive axes. Thus, one can obtain ideal networks with three (octahedron), four (tetrahedron and cube), six (icosahedron), and ten (dodecahedron) sensors through this method; denoted $N_3$, $N_4$, $N_6$, and $N_{10}$. These arrangements have the sensitivity $\minMag=\frac{\sigma/\kappa}{\sqrt{n/3}}$ and are irreducible. With these networks alone, it is evident that there are multiple unique ways of orienting a given number of sensors that do not rely on using the same set of ideal subnetworks; for example, six sensors can be arranged as $N_6$ or $N_3 \oplus N_3$. Further, these networks can be combined to form ideal networks with three, four, six, or more sensors. 

What seems to remain is a way to orient five identical, independent sensors into an ideal network. One can show that two such network arrangements $N_{5a}$ and $N_{5b}$ are given by 
\begin{widetext}
\begin{equation}
D_{5a} = \kappa \left[ \begin{matrix}
0 &  \sqrt{\frac{5}{6}} & \sqrt{\frac{1}{6}} \\
0 & -\sqrt{\frac{5}{6}} & \sqrt{\frac{1}{6}} \\
 \sqrt{\frac{5}{6}} & 0 & \sqrt{\frac{1}{6}} \\
-\sqrt{\frac{5}{6}} & 0 & \sqrt{\frac{1}{6}} \\
0 & 0 & 1
\end{matrix} \right]\quad\text{and}\quad
D_{5b} = \kappa \left[ \begin{matrix}
1 & 0 & 0 \\
-\sqrt{\frac{1}{3}} & -\sqrt{\frac{2}{3}} & 0 \\
-\sqrt{\frac{1}{3}} &  \sqrt{\frac{2}{3}} & 0 \\
0 &  \sqrt{\frac{1}{6}} & \sqrt{\frac{5}{6}} \\
0 & -\sqrt{\frac{1}{6}} & \sqrt{\frac{5}{6}}
\end{matrix} \right]\,,
\end{equation}
\end{widetext}
each with sensitivity $\minMag = \sqrt{\sigma/\kappa}{\sqrt{5/3}}$. These networks are unique, even when considering parity reversal of individual sensors and global rotations; this is evident because $D_{5b}$ has orthogonal sensor while $D_{5a}$ does not. Further, these are irreducible ideal networks because $n<6$. Along with $N_3$ and $N_4$, one can generate any ideal network with at least three identical, independent sensors using these arrangements. 

Though one can always arrange three or more identical, independent sensors into an ideal network, it is not always the case that a given set of sensors can be arranged into an ideal network. For example, consider a set of $n$ sensors all with the same coupling $\kappa=1$, but $n-1$ magnetometers have noise $\sigma_0$ and the last magnetometer has noise $\sigma_1 < \sigma_0/\sqrt{(n-1)/2}$. An optimized network would be arranged with the first $n-1$ sensors evenly oriented around the plane orthogonal to the last sensor and have a quality $\quality=\frac{\sigma_1}{\sigma_0}\sqrt{\frac{n-1}{2}} < 1$ and sensitivity $\minMag_0=\sigma_1$. The least-sensitive direction is orthogonal to the last sensor's sensitive axis and any adjustments to the sensors' orientations would worsen the sensitivity in this plane. The problem in this scenario is that one sensor is much more sensitive than the others, so they cannot compensate, even collectively. The extreme case of this would be if the first $n-1$ sensors were so noisy that they were effectively inoperable; this arrangement would not be much different than an $n=1$ network, which cannot be ideal.

\subsection{Optimizing networks}
Regardless of whether a given set of magnetometers can be arranged into an ideal network, their arrangement can always be optimized; at least at a given time. In practice, the noise in each GNOME magnetometer varies over time and magnetometers turn on and off throughout the experiment. 

A non-optimized network can still be improved via an algorithm. An example of a ``greedy'' algorithm would be one in which, each step, a sensor is randomly selected, removed from the network, and re-inserted in the least-sensitive direction for the network without the removed sensor. This step can then be repeated many times until reaching some optimization condition. 
For multi-axis sensors that always have the same relative angle between the sensitive axes, the orientation by which to re-insert the sensor is a bit more complicated. Roughly, one would apply a rotation to the sensor to align the best- and worst-directions for the multi-axis sensor and the rest of the network (see Appendix~\ref{app:multiAxis}).

This algorithm will also work regardless of whether an ideal network arrangement exists, though it is not possible to tell if the resulting network is truly optimal. 

\begin{table}
    \centering
    \caption{Optimizing GNOME for each run. In particular, a network is constructed using all magnetometers active for at least 25\,\% of the run. Noise is calculated as the average standard deviation of the data $\left< \sigma^{-1} \right>^{-1}$ after applying a 1.67\,mHz high-pass filter, 20\,s averaging, and notch filters to remove powerline frequencies. The columns contain the run number, number of sensors, network characteristics, optimized network characteristics, theoretical optimized sensitivity from Eq.~\eqref{eq:optSensitive}, and factor by which the optimization algorithm improved sensitivity.}
    \begin{tabular}{ll ll ll ll}
        \hline\hline
        & & \multicolumn{2}{c}{During run} & \multicolumn{2}{c}{Optimized} & & \\
        Run & Size & $\minMag_0$ (pT) & $\quality$ & $\minMag_0$ (pT) & $\quality$ & $\minMag_\text{opt}$ (pT) & Improved \\
        \hline
            1 & 6 & 0.74 & 0.19 & 0.27 & 0.62 & 0.22 & 2.72 \\
            2 & 9 & 0.48 & 0.49 & 0.30 & 1    & 0.30 & 1.59 \\
            3 & 7 & 0.51 & 0.27 & 0.26 & 0.61 & 0.21 & 1.97 \\
            4 & 9 & 1.13 & 0.22 & 0.42 & 0.79 & 0.37 & 2.67 \\
            (pre)5 & 9 & 0.25 & 0.51 & 0.17 & 0.82 & 0.16 & 1.40 \\
        \hline\hline
    \end{tabular}
    \label{tab:runOptimized}
\end{table}

The optimization can be applied to the GNOME network to better understand the potential of the experiment. In particular, this was performed for GNOME's official Science~Runs:
\begin{itemize}
    \item Science Run~1: 6 June--5 July 2017.
    \item Science Run~2: 29 November--22 December 2017.
    \item Science Run~3: 1 June 2018--10 May 2019.
    \item Science Run~4: 30 January--30 April 2020.
    \item Science pre-Run~5: 16 June--23 August 2021.
\end{itemize}
The results of the optimization is given in Table~\ref{tab:runOptimized} in which the coupling was assumed to be one. For most of the Science Runs, the improvement was less than a factor of two with the largest improvement being by a factor of 2.7 from Science Run~4. 
Additionally, the Science Run~2 network could be made ideal with about the same sensitivity as predicted in Eq.~\eqref{eq:optSensitive}.
It should be noted that the networks used in the optimization included most of the sensors available, while it is not typical for all the sensors in GNOME to be active at the same time. 
Another example of network optimization is given in Appendix~\ref{app:example} wherein only a couple sensors are reoriented and an additional sensor is added.

\section{Conclusions}\label{sec:conclusion}
In this paper, we have considered a network of sensors with directional sensitivity and how the collective sensitivity of the network is affected by the choice in how the sensors are oriented. To do this, a ``quality factor'' was introduced to quantify how efficiently the network sensors are oriented regardless of their underlying sensitivity. Various properties and examples of ideal networks were presented, along with a means of optimizing an existing network. By optimizing GNOME, we can show some modest improvement in the network sensitivity without requiring additional sensors or improvements in existing sensors. 

In practice, there are still some limitations in how sensors can be oriented. For example, depending on how the sensor is built, it may not be possible/practical to reorient it, or it may only be possible to orient its sensitive axis vertically or horizontally. Additionally, what may be an optimal network under some set of conditions, it is difficult to predict how the sensitivities of the sensors will change and which ones will be active. However, if there are many decent/reliable sensors, the network as a whole will not usually deviate too much from its optimal arrangement. 

This work only considered the manner in which the sensors were oriented, not their position. The relative positions and distances between the sensors is not relevant to understanding the sensitivity of the network as a whole. Briefly, the optimal placement of the sensors in a network differs depending on the goal of the experiment. It is generally simpler to observe a signal that crosses a network of nearby sensors; because the potential crossing time is shorter, less data need to be compared between the sensors. However, measuring the direction and speed of some phenomena crossing the network can be done more accurately when the sensors are more distant from one another. 

The work described in this paper can be applied to any experiment involving a network of directionally sensitive devices. These networks are useful in detecting features in a vector field or gradient that traverse a spatial region. Both the design and improvement of these networks can be meaningfully improved through careful consideration of how sensors are oriented.

Moving forward, this study can have a direct influence on Advanced GNOME; a planned general upgrade to the GNOME experiment. This upgrade includes the addition of SERF comagnetometers~\cite{shah_fully_2018} with the option to operate in two-axis mode. The use of multi-axis sensors can be incorporated into the work presented here by treating them as multiple sensors with correlated noise. These sensors also have the constraint that the sensitive axes must remain orthogonal. Optimizing networks with these additional constraints and complexities will be interesting future work.

\section*{Acknowledgments}
Work for this paper was made possible through discussions with the GNOME collaboration as well as characterization data from the experiment. Discussions with  Dr.~Derek Jackson Kimball, Dr.~Dmitry Budker, Dr.~Szymon Pustelny, Dr.~Ibrahim Sulai, and Hector Masia Roig greatly aided in the completion of this work. 
JAS has no funding sources to declare.

\bibliographystyle{naturemag}
\bibliography{netQualBib}

\appendix

\section{Multi-axis sensors} \label{app:multiAxis}
Consider a multi-axis magnetometer that can only be reoriented by rotating the entire sensor --- one cannot adjust individual sensitive axes. Some effort is needed to understand how to orient such a sensor with respect to a larger network.

Let $N_a=\{D_a,\Sigma_a\}$ describe the orientation of the multi-axis sensor and $N_b=\{D_b,\Sigma_b\}$ describe the rest of the network. The matrices $D_i^T\Sigma D_i$ (for $i\in\{a,b\}$) can be diagonalized as $U_i\Lambda_i U_i^T$ where the $\Lambda_i$ is a diagonal matrix whose $j^\text{th}$ element along the diagonal is the $j^\text{th}$-largest eigenvalue $\lambda_ij$, and $U_i$ is an orthogonal matrix ($U_i^{-1}=U_i^T$) whose $j^\text{th}$ column is the respective (normalized) eigenvector $\boldsymbol{u}_{ij}$. The orthogonal matrices have the effect of rotating the coordinate axes to the best- and worst-directions (i.e., most- and least-sensitive) for the respective network: $\boldsymbol{\hat{x}}\mapsto$ the worst-direction and $\boldsymbol{\hat{x}}\mapsto$ the best-direction. Finally, define $\tilde{U}_i$ as $U_i$ with its columns reversed; this also reverses which coordinate axis will be rotated to which direction.

When orienting the multi-axis sensor to be included in the rest of the network, it is optimal to orient the respective best-direction of the multi-axis sensor with the worst-direction of the rest of the network and vice-versa. This is accomplished by rotating the multi-axis sensor as follows:
\begin{equation}
    D_a \to D_a \left( \tilde{U}_b U_a^T \right)^T\,.
\end{equation}
This rotation maps:
\begin{align*}
    \tilde{U}_b U_a^T:& \text{worst for}\ a\mapsto \boldsymbol{\hat{x}} \mapsto \text{best for}\ b \\
    & \text{best for}\ a\mapsto \boldsymbol{\hat{z}} \mapsto \text{worst for}\ b\,.
\end{align*}

For example, the Hayward and Krakow sensors could operate as two-orthogonal-axis magnetometers during Science pre-Run~5; though at the cost of worse sensitivity, roughly doubling the variance. Replacing these two sensors during the Science Run with (uncorrelated) two-axis sensors and including them with optimal orientations improves the sensitivity by a factor of~1.17 to $\minMag_0 = 0.21\,\text{pT}$. This is slightly better than the 1.15~factor of improvement for optimizing the two sensors in single-axis mode.

\section{Applied optimization} \label{app:example}
It is useful to consider an explicit examples of optimizing parts of the network. Specifically, consider reorienting the Mainz and Krakow magnetometers and adding another magnetometer with 0.2\,pT standard deviation noise after filtering/averaging. The new magnetometer will be located in Berkeley, CA, USA. This optimization will use the Science pre-Run~5 characteristics.

Locally, the orientation is expressed in horizontal coordinates; using altitude (alt) and azimuthal (az) angles expressed as the pair (alt, az). The altitude is the angle relative to the horizon, while the azimuth is the angle relative to noise (measured clockwise). The Mainz sensor is oriented with $\text{alt}=-90^\circ$ while the Krakow sensor is oriented as $(50^\circ, 20^\circ)$ during Science pre-Run~5. 

Optimizing the direction of the three magnetometers would result in a sensitivity of $\minMag_0=0.16\,\text{pT}$ with $\quality=0.90$. Using Eq.~\eqref{eq:optSensitive}, the optimal sensitivity would be $\minMag_\text{opt} = 0.15\,\text{pT}$. When optimized, the Mainz magnetometer is oriented as $(58^\circ, 81^\circ)$, the Krakow magnetometer is oriented as $(13^\circ, -31^\circ)$, and the Berkeley magnetometer is oriented as $(4^\circ, -87^\circ)$.


\end{document}